\documentstyle[12pt]{article}
\newcommand{\sect}[1]{\setcounter{equation}{0}\section{#1}}

\newcommand{\be}{\begin{equation}}
\newcommand{\en}{\end{equation}}
\newcommand{\bea}{\begin{eqnarray}}
\newcommand{\ena}{\end{eqnarray}}
\newcommand{\vs}[1]{\vspace{#1 mm}}
\newcommand{\ii}{r}
\newcommand{\sm}[2]{\frac{\mbox{\footnotesize #1}\vs{-2}}
                   {\vs{-2}\mbox{\footnotesize #2}}}
\newcommand{\PL}{Phys. Lett. }
\newcommand{\W}{\mbox{\sf W}}
\begin{document}
%
%
%
%
%
%
%
%
%
%
\renewcommand{\thefootnote}{\fnsymbol{footnote}}
\newpage
\setcounter{page}{0}
\pagestyle{empty}
\leftline{KCL-TH-92-1}
\leftline{June 1992}

\vs{20}

\begin{center}
{\LARGE {Commuting quantities and exceptional \W-algebras}}\\[1cm]
{\large M.D. Freeman, K.\ Hornfeck and P. West}\\[0.5cm]
{\em King's
College, Department of Mathematics, Strand,
London WC2R 2LS, GB}
\\[1cm]
\end{center}
\vs{15}

\centerline{ \bf{Abstract}}
Sets of commuting charges constructed from the current of a \mbox{\sf U}(1)
Kac-Moody algebra are found.  There exists a set $S_n$ of such charges for each
positive integer $n>1$; the corresponding value of the central charge in the
Feigin-Fuchs realization of the stress tensor is $c=13-6n-6/n$.  The charges in
each series can be written in terms of the generators of an exceptional
\W-algebra.
\renewcommand{\thefootnote}{\arabic{footnote}}
\setcounter{footnote}{0}
\newpage
\pagestyle{plain}

%
%
%
\sect{Introduction}
Classical integrable systems have been studied
for many years.  Two of the most familiar examples are the KP hierarchy,
with its reductions to the KdV and higher hierarchies (and also the modified
KdV hierarchies), and affine Toda
field theory, of which the simplest is sinh-Gordon theory.  The
integrability of these systems stems from the presence of an infinite
number of conserved quantities that commute with each other.  The existence
of these conserved quantities can be shown in a number of different ways,
for example by using pseudodifferential operators, or by Hamiltonian
reduction from affine Kac-Moody algebras.  There is in fact a link between the
KdV, mKdV and sinh-Gordon equations, coming from the fact that, after a change
of variables, the conserved quantities for the KdV equation coincide with
those of the mKdV and sinh-Gordon equations.  There are analogous links for the
other hierarchies of integrable systems referred to above.

The KdV, mKdV and sinh-Gordon equations can all be written in Hamiltonian
form, and in fact for the KdV equation there is more than one way to do this.
There is an intruiging connection between the so-called second Poisson bracket
structure of the KdV equation and the Virasoro algebra \cite{Gervais},
satisfied by the $T_{zz}$-component of the energy-momentum tensor in conformal
field theories.  In fact, the components of the field variable $u$ of the KdV
equation have Poisson brackets given precisely by the Virasoro algebra.

More recently the quantum analogues of the above integrable systems have
been investigated.  A significant contribution was made by Sasaki and
Yamanaka \cite{SY}, who investigated the quantum Sine-Gordon
equation and its relation with the quantum KdV (qKdV) and quantum mKdV
equations. They found that at low levels the sine-Gordon equation did indeed
possess conserved quantities, and they  suggested that conserved quantities
would exist at all levels.  Furthermore, they realized that these conserved
quantities could be expressed in terms of the energy-momentum tensor associated
with the related qKdV equation.

In another development Zamolodchikov \cite{Z} showed that certain perturbations
of conformal field theories lead to conserved quantities in addition to those
associated with the two-dimensional Poincar\'e group.  These additional
quantities were certain polynomials in the energy-momentum tensor, and were
shown to exist at low levels by an elegant counting argument.  Furthermore, by
explicit calculation it was shown that the first two non-trivial charges
commuted.  It was suggested that there were in fact an infinite number of
conserved quantities, and that they commuted amongst themselves.  Affine Toda
field theories
were studied as deformations of conformal field theories by Eguchi and Yang
\cite{EY1} and by
Hollowood and Mansfield \cite{HollowoodMansfield}, generalizing the work of
ref.  \cite{SY}, and Kupershmidt and Mathieu \cite{KM} considered the qKdV
equation in the context of deformed conformal field theories.

Unfortunately the arguments that can be used to determine the form and
properties of the conserved quantities in the classical theory do not seem to
have straightforward analogues in the quantum domain. One can, for example,
write the qKdV equation in terms of
a Lax pair involving the operator $T$, but the absence of a natural associative
normal-ordered product for differential polynomials in $T$ means that this
Lax pair cannnot be used to prove the existence of infinitely many
commuting charges.  To show the commutativity of even the low level conserved
quantities requires a lengthy calculation, and although the first few levels
have been checked by computer the existence of these quantities appears
somewhat miraculous.  Recently Feigin and Frenkel \cite{FeiginFrenkel} have
shown the existence of an infinite set of commuting quantities in a free field
realization, using homology arguments, but this proof does not give explicit
forms for the conserved charges.

Some partial results are known, however.  For the
central charge $c$ having the value $-2$, the general form
of the conserved quantities was conjectured
by Sasaki and Yamanaka \cite{SY} and recently proven by DiFrancesco, Mathieu
and Senechal \cite{DiFMS} using a free fermion realisation of the $c=-2$
Virasoro algebra.  Also, for the $(p,p') = (2,2k+1)$ series
of minimal models, the density of the charge with spin
$2k-1$ is equal to the singular vector that occurs in the
vacuum representation of the Virasoro algebra at level
$2k$ for $c=1-3(2k-1)^2/(2k+1)$ \cite{FKM,EY2,KNS,DiFM}.

Clearly, one of the outstanding problems in this development is to gain a
better understanding of the conserved quantities and their properties.  One way
in which this might be achieved would be if the conserved quantities were
contained in some larger algebra within which they could be viewed as some sort
of Cartan subalgebra.
There are two situations in which this picture does indeed arise.  The
first is the case of $c=-2$.  In this case the enveloping algebra of the
Virasoro algebra contains a truncation of \W$_\infty$ \cite{PRS1,PRS2} as a
linear subalgebra, and it is easy
to see that this subalgebra contains a `wedge subalgebra' \cite{PRS2} which
includes
infinitely many commuting charges.  In fact these charges are simply the
integrals of the basic even-spin quasiprimary fields $V^i$ of \W$_\infty$.  The
second
situation in which the picture we are thinking of arises is the classical
case, viewed from the perspective of $\tau$-functions.  Here the commuting
quantities of the KdV equation can be embedded in the affine Lie algebra
$\widehat{\mbox{\sf SU}(2)}$ \cite{DKJM}, and in fact they are  simply the
elements $J^3_n$, $n>0$ of this Kac-Moody Lie algebra.

In this paper we attempt to gain some insight into these issues by looking for
commuting charges that can be constructed from the generators of a \mbox{\sf
U}$(1)$ Kac-Moody algebra. We began our investigation by using a computer to
look for commuting charges, and we then interpreted our results analytically.
Since the Virasoro algebra has a free field realization we expected to obtain
the charges corresponding to the qKdV equation, but we found in addition that
it is possible to construct extra commuting quantities.  In particular we found
that there exists an infinite number of series of commuting charges, with each
series itself containing an infinite number of charges.  In terms of the free
field realization of the Virasoro algebra each series exists only for a
particular value of $c$.  The $n$'th series contains even spin currents for
every even value of the spin, but in addition contains odd spin currents at
spins $m(2n-2)+1$ for $m$ a positive integer.  The lowest such non-trivial odd
spin is $2n-1$, and the corresponding current can be taken to be a primary
field. Furthermore this field and the identity operator generate a \W-algebra,
denoted by \W$(2n-1)$, which plays an important role in determining the
structure of the charges. We were able to prove analytically that these charges
did commute, and furthermore we found that the \W-algebra and an extension of
it provide a natural algebraic framework in which this result can be
understood.  Eguchi and Yang \cite{EY2}, in their work on the restricted
sine-Gordon equation \cite{LC,S}, noted that for the values of $c$ we
considered there should exist additional even-spin charges.  However, the forms
of the charges they suggest differ from ours, and they did not consider any
connection with an underlying \W-algebra.

\sect{Computer results}
We consider the \mbox{\sf U}$(1)$ Kac-Moody algebra $\widehat{\mbox{\sf
U}(1)}$,
\be
[J_n,J_m] = n \delta_{n+m,0},
\label{KM}
\en
with generators $J_n = \oint dz\, z^n J(z)$ and
corresponding operator product expansion
\be
J(z) J(w) = {1 \over (z-w)^2}+ \ldots\quad.
\label{JJ}
\en
We wish to address the question of what sets of mutually
commuting quantities can be constructed as integrals of
polynomials in $J$ and its derivatives.  To put this another
way, we want to find sets of mutually commuting operators in
the enveloping algebra of $\widehat {\mbox{\sf U}(1)}$. We know that it
is possible to write a stress energy tensor in terms of $J$,
\be
T=\sm{1}{2} :J^2: + \alpha J',
\label{FF}
\en
and so we certainly expect to find the commuting charges
that correspond to the KdV equation.  We
shall therefore look for sets of mutually commuting charges
that include operators not contained in this series of charges.

In this section we report on our investigations of the above
question using a computer to carry out the operator product
expansions.  To do this we used Mathematica \cite{W} and the operator
product package OPEdefs \cite{T}.  As a consequence of doing the
algebra by computer, however, these results are necessarily
restricted to currents of relatively low spin; we have
looked at currents of spin 13 and less.  In subsequent
sections we give analytic derivations of many of these
results, thereby extending them to arbitrary spins.

Given currents $A$ and $B$ with charges
\be
Q_A=\oint dz\,A(z),\quad Q_B = \oint dz\,B(z),
\en
the commutator of $Q_A$ with $Q_B$ is given by
\be
[Q_A,Q_B] = \oint_0 dw \oint_w dz\, A(z) B(w).
\label{commutator}
\en
The only contribution to this commutator comes from the
single pole term in the OPE of $A$ and $B$, so
$Q_A$ and $Q_B$ will commute precisely when this single pole
term is a derivative.

It is straightforward to see that
\be
Q_1 \equiv \oint dz\, J(z)
\label{Q_1}
\en
commutes with any charge constructed from $J$ and its
derivatives.  It is also true that
\be
Q_2 \equiv \oint dz\, :\!J^{\,2} (z)\!:
\label{Q_2}
\en
commutes with all other charges, since if $P$ is an
arbitrary differential polynomial in $J$, the single pole
term in $:\!J^2(z)\!: P(w)$ comes from the single contraction
term, namely
\be
2 \sum_{n=0}^\infty{(n+1)! \over (z-w)^{n+2}} :\!J(z)\,
{\partial P(w) \over \partial J^{\,(n)}(w)}\!:\ ,
\en
and the coefficient of the single pole in this expression is
just
\be
2 \sum_{n=0}^\infty :\!J^{\,(n+1)}\, {\partial P \over \partial
J^{\,(n)}}\!:\ \  =2 P'
\en
This is also apparent from the Feigin-Fuchs
representation---the integral of $1/2 :\!\!J^{\,2}\!\!:$ is just $L_{-1}$,
the generator of translations.

In order to go beyond these somewhat trivial commuting
charges, we looked for charges commuting with the
integral $Q_4$ of the spin-4 current
\be
p_4 = \,:\!J^{\,4}\!: +\, g :\!J'^{\,2}\!:
\label{p_4}
\en
for some value of $g$, using a computer to calculate the OPE's.
We found that there were a number of different sets of such
charges.  All of the currents in a given set gave rise to
charges that commuted with a given $Q_4$, with the same
value of the coupling $g$ for each set.  Each set  contained
a unique current at every even spin, modulo total
derivatives, corresponding to those of the qKdV equation.
There were also currents occuring at certain odd spins.
In fact each set was uniquely fixed by the spin of the
lowest non-trivial odd spin current.  If the spin of this
current is $h\equiv 2n-1$, we call the corresponding set of currents
$S_n$.  The odd spin currents in $S_n$ then occur at spins
$1,h, 2h-1, 3h-2,\ldots$, increasing in steps of $h-1$; the
charges constructed from these currents commute with $Q_4$
provided that $g$ is given in terms of $h$ by
\be
g={1\over h+1} (h^2-4h-1)
\label{g,h}
\en
We found such sets of odd spin currents for every odd
integer $h$ larger than $1$.  Thus the first series $S_2$ has $h=3$ and
has a unique current at every odd spin as well as at every
even spin.  The second series $S_3$ has unique odd spin
currents at spins $5,9,13,\ldots$ as well as currents at
every even spin.

Given a set of charges commuting with a given spin-3 charge
$Q_4$, it is natural to ask whether these charges commute
with each other.  We have verified that this is indeed the
case for the currents we found using the computer, and we
shall give a general analytic proof in section 4.

Although it would clearly be possible to give expressions for the currents
series by series, it is remarkable that there is a relatively simple
formula that gives all the currents in all the series, at
least up to the dimension to which we have calculated.  In order to write
down such a formula we include a parameter $x$ that is given in
terms of the coupling constant $g$ by $x=-3(g+1)$.  The
reason for this choice of parametrization is that for $x=0$
the currents take particularly simple forms; this will be
explained fully in a later section.  Since the currents
are fixed only up to total derivatives it is useful to
make a choice of basis for the space of fields modulo
derivatives.  Making such a choice, the
expression for $p_r$ is
\bea
p_\ii & = & :\!J^{\,\ii}\!: + :\!J^{\,\ii-4} J'^{\,2}\!: g_1(\ii,x) \, +
:\!J^{\,\ii-6}J''^{\,2}\!: g_2(\ii,x)\, +
\nonumber \\[1mm]
&&:\!J^{\,\ii-8}J'''^{\,2}\!: g_3(\ii,x)\, +  :\!J^{\,\ii-8}J'^{\,4}\!:
g_4(\ii,x)\,
+ :\!J^{\,\ii-9}J''^{\,3}\!: g_5(\ii,x) +
\nonumber \\[1mm]
&&:\!J^{\,\ii-10}{J^{\,(4)}}^2\!:\, g_6(\ii,x) \,+
:\!J^{\,\ii-10}J'^{\,2}J''^{\,2}\!: g_7(\ii,x) +  \ldots
\label{fields}
\ena
where the coefficients $g_l(\ii,x)$ are polynomials in the
exansion parameter $x$.  The coefficients
$g_1(\ii,x),\ldots,g_5(\ii,x)$ are given by
\bea
g_1(\ii,x) & = &- \frac{1}{3} \left(\begin{array}{c} \ii \\
4
\end{array} \right)
\frac{1}{(\ii-3)} \left[ 3\,(\ii-3) + x \right] \nonumber
\\[1mm]
g_2(\ii,x) & = & \frac{1}{6} \left(\begin{array}{c} \ii \\ 6
\end{array} \right)
\frac{1}{(\ii-3)\,(\ii-5)} \left[ 6\,(\ii-3)(\ii-5) + 5\, x
\,(\ii-4) + x^2
\right] \\[1mm]
g_3(\ii,x) & = & -\frac{1}{9} \left(\begin{array}{c} \ii \\
8
\end{array} \right)
\frac{1}{(\ii-3)\,(\ii-5)\,(\ii-7)}\,\mbox{\small $\times$}\nonumber\\[1mm]
&&\hspace{-1cm}\left[ 9\,(\ii-3)(\ii-5)(\ii-7) + \sm{7}{10}\, x \,
(19\ii^2-192\ii+450) \,+ \sm{1}{30}\, x^2 \,(193\ii-962) + x^3
\right]\nonumber\\[1mm]
g_4(\ii,x) & = & - \frac{7}{27} \left(\begin{array}{c} \ii
\\
8 \end{array} \right)
\frac{1}{(\ii-3)\,(\ii-5)\,(\ii-7)}\,\mbox{\small $\times$}\nonumber\\[1mm]
&&\hspace{-1cm}\left[ 27\,(\ii-3)(\ii-5)(\ii-7) + 6\, x \,
(4\ii^2-42\ii+105)\, + x^2 \,(5\ii-31) \right]\nonumber
\\[1mm]
g_5(\ii,x) & = & \frac{10}{9} \left(\begin{array}{c} \ii \\
9 \end{array} \right)
\frac{1}{(\ii-3)\,(\ii-5)\,(\ii-7)}\,\mbox{\small $\times$}\nonumber\\
&&\hspace{-1.5cm}\left[ 9\,(\ii-3)(\ii-5)(\ii-7) + \sm{63}{100} \,x \,
(21\ii^2-213\ii+500) \,+\sm{1}{100}\, x^2\, (641\ii-3209) + x^3
\right]\nonumber.
\label{g}
\ena
These coefficients describe the fields $p_r$ up to
$\ii = 9$ exactly and give the first terms for all the
higher dimensional fields. For completeness we list explicitly the additional
terms in the fields of spin 10 to 13 that are not described by the coefficients
$g_1,\ldots,g_5$:
\bea
p_{10} &=& \ldots +\left( 90 + {152\over3} x + {2381\over 315} x^2 +
{17\over 63} x^3 \right)  {J'}^2 {J''}^2\nonumber\\[1mm]
&& + \left( 1+ {565\over 756} x + {36311\over 158760} {x^2} +
{178\over 6615} {x^3} + {1\over 1134} {x^4} \right)
       {{J^{(4)}}^2}\nonumber\\[2mm]
p_{11} &=& \ldots + 11\left( 90 + {6263\over 192} x + {689\over 192} {x^2} +
{5\over 48} {x^3} \right) J {{J'}^2}
       {{J''}^2}\nonumber\\[1mm]
&& -11 \left( 5 + {26329\over10752}x + {12407\over24192} {x^2} +
{13135\over290304} {x^3} + {25\over 20736} {x^4} \right)  J''
{{J^{(3)}}^2}\nonumber\\[1mm]
&&+11 \left( 1 + {16519\over32256} x + {643\over6048} {x^2} +
{2663\over 290304} {x^3} + {5\over20736} {x^4}\right) J
{{J^{(4)}}^2}\nonumber\\[2mm]
p_{12} &=& \ldots    -11 \left( 27 + {251\over 25} x + {14939\over 14175} {x^2}
+ {79\over2835} {x^3} \right)
        {{J'}^6}\nonumber\\[1mm]
&& + 22  \left( 270 + {378\over5} x + {6319\over945} {x^2} + {31\over 189}
{x^3} \right)
        J^2{{J'}^2} {{J''}^2}\nonumber\\[1mm]
&& + 11 \left( 20 + {1228\over135} x + {6929\over3969} {x^2} +
{8173\over59535} {x^3} + {11\over 3402} {x^4} \right)
{{J''}^4}\nonumber\\[1mm]
&& - 11 \left( 15 +{478\over45} x + {100301\over 39690} {x^2} +
{4309\over19845} {x^3} + {1\over89} {x^4}
          \right)  {{J'}^2} {{J^{(3)}}^2}\nonumber\\[1mm]
&& - 11  \left( 60 + {1054\over45} x + {450913\over119070} {x^2} +
{3151\over11907} {x^3} + {10\over 1701} {x^4} \right) J J''
{{J^{(3)}}^2}\nonumber\\[1mm]
&& + 11  \left( 6 + {109\over45}x + {92891\over238140} {x^2} + {319\over11907}
{x^3} + {1\over 1701} {x^4} \right) J^2 {{J^{(4)}}^2}\nonumber\\[1mm]
&& - \left( 1 + {7601\over9450} x + {6457273\over21432600} {x^2} +
{294211\over5358150} {x^3} + {4511\over1071630} {x^4} + {1\over1020600} {x^5}
\right)  {{J^{(5)}}^2}\nonumber\\[2mm]
p_{13} &=& \ldots  -143  \left( 27 + {{3623}\over {480}}x +
   {{4579}\over {7200}}x^2 + {{31}\over {2160}}x^3\right)
       J {{J'}^6}\nonumber\\[1mm]
&&+ 143   \left(180 + {{4003}\over {96}}x +
   {{1117}\over {360}}x^2 + {{19}\over {288}}x^3\right)
       J^3 {{J'}^2} {{J''}^2}\nonumber\\[1mm]
&& + 143 \left( 30 + {{114091}\over {8064}}x +
   {{76861}\over {30240}}x^2 +
   {{343}\over {1920}}x^3 + {{97}\over {25920}}x^4\right)
   {{J'}^2} {{J''}^3}\nonumber\\[1mm]
&& + 143  \left( 20 + {{57319}\over {8064}}x +
   {{252527}\over {241920}}x^2 +
   {{283}\over {4320}}x^3 + {{1}\over {768}}x^4\right) J
{{J''}^4}\nonumber\\[1mm]
&& - 143  \left( 15 + {{28435}\over {4032}}x +
   {{29731}\over {24192}} x^2+
   {{1453}\over {17280}}x^3 + {1\over {576}}x^4\right)J {{J'}^2}
{{J^{(3)}}^2}\nonumber\\[1mm]
&& - 143   \left(30 + {{158489}\over {16128}}x +
   {{9883}\over {7560}}x^2 +
   {{175}\over {2304}}x^3 + {{5}\over {3456}}x^4\right) J^2 J''
{{J^{(3)}}^2}\nonumber\\[1mm]
&& + 143  \left( 2 + {{32587}\over {48384}}x +
   {{3245}\over {36288}}x^2 +
   {{59}\over {11520}} x^3+ {1\over {10368}}x^4\right) J^3
{{J^{(4)}}^2}+\nonumber\\[1mm]
&&\hspace{-5pt}  13 \left( 7 + {{417901}\over {107520}}x +
   {{14787253}\over {14515200}}x^2 +
   {{6047609}\over {43545600}}x^3 +
   {{53143}\over {6220800}}x^4 +
   {{7}x^5\over {41472}}\right)  J'' {{J^{(4)}}^2}\nonumber\\[1mm]
&&\hspace{-5pt} - 13  \left( 1 + {{57013}\over {96768}}x +
   {{83483}\over {537600}}x^2 +
   {{17987}\over {870912}}x^3 +
   {{1541}\over {1244160}}x^4 + {x^5\over {41472}}\right)
   J {{J^{(5)}}^2}.\nonumber
\ena

We shall see that the existence of the infinitely many sets
of currents $S_n$ is related to the
occurence of certain \W-algebras.  The first step in this
process is the verification that the even spin currents can
be expressed in terms of the Feigin-Fuchs energy-momentum
tensor given in eqn (\ref{FF}), which has background charge
$\alpha$ and central charge $c= 1-12 \alpha^2$.
Evidently $Q_2 = \oint \!:\!J^{\,2}\!: \,= 2 L_{-1}$.  The charge $Q_4$
can also be written in terms of $T(z)$, provided we relate
the coupling $g$ and the central charge $c$ by
\be
g=-\sm{1}{3} (c+5);
\label{g,c}
\en
for this value of $g$ we have
\bea
Q_4 & \equiv & \oint :\!J^{\,4}\!: +\, g\, :\!J'^{\,2}\!:\nonumber \\
    & = & 4 \oint :\!T^2\!:\quad.
\label{TJ}
\ena
Here $T^2$ is normal-ordered according to the standard prescription in
conformal field theory, which takes the normal product of two
operators $A(z)$ and $B(w)$ to be the term of order $(z-w)^0$ in their
OPE \cite{BBSS},
\be
(AB)(Z) \equiv \oint \, dw { A(w) B(z) \over w-z }.
\label{NOP}
\en
To reconcile the two expressions for $Q_4$ in (\ref{TJ}), we must relate
the normal
ordering in terms of $J$, which is the usual Wick
ordering, to that in terms of $T$.  Using the formula \cite{BBSS}
\be
((AB)C) = A(BC) + [(AB),C] - A[B,C] - [A,C]B
\en
with $A=B=J$ and $C=\,:\!J^2\!:$, and also making use of
\be
:\!J:\!J^2\!:: - ::\!J^2\!:J \!:\,= -J'',
\en
it is straightforward to recover eqn(\ref{TJ}).

In this way it is possible to express the even spin currents in
terms of $T$ alone, provided we use the freedom to add and
subtract total derivatives, but the odd spin currents cannot be expressed only
in terms of $T$.

The value $c_n$ of the central charge corresponding to the
series $S_n$ can be found from equations (\ref{g,h}) and (\ref{g,c})
to be
\be
c_n = 13 -6n -{6\over n}.
\label{cn}
\en
It is known for $n=2$, 3 and 4 that a \W-algebra \W$(2n-1)$ exists \cite{KW,B},
where \W$(2n-1)$ is an algebra generated by the identity operator together with
a single primary field of spin $2n-1$.  For $n=2$ this algebra is the
\W$(3)$-algebra of Zamolodchikov \cite{Z2}, which exists for any value of $c$,
while for $n=3$ the algebra \W$(5)$ exists for only five values of $c$, namely
6/7, $-350/11$, $-7$, $134\pm 60/\sqrt5$, and for $n=4$ \W$(7)$ exists for the
single value $-25/2$ of $c$.  There is an argument to suggest that an algebra
\W$(2n-1)$ exists for any $n$ at the value of $c$ given in eqn (\ref{cn}).
This can be seen by examining the primary field algebra for the conformal field
theory with $c$ given by (\ref{cn}).  In this model the field $\phi_{(3,1)}$
has dimension $2n-1$ and is expected to have an OPE with itself of the form
\cite{BPZ}
\be
[\phi_{(3,1)}][\phi_{(3,1)}] = [1] + [\phi_{(3,1)}] + [\phi_{(5,1)}].
\label{3,1,3,1}
\en
The dimension of $[\phi_{(5,1)}]$ is $6n -2$, however, which is too high to
appear in the OPE of a spin $2n-1$ field with itself, and so we expect the
identity operator together with $[\phi_{(3,1)}]$ to form a closed operator
algebra.   This observation has also been made previously by Kausch \cite{K}.
We note in addition that for the values of $c$ we are considering the field
$[\phi_{(3,1)}]$ has odd spin, and as a consequence cannot appear on the
right-hand side of the OPE (\ref{3,1,3,1}).  Thus the OPE of $[\phi_{(3,1)}]$
with itself gives only the identity operator in this case.

It is natural to
suppose that the series $S_n$ is associated in some way with
the algebra \W$(2n-1)$.  This is confirmed by the realization that
the first non-trivial field of odd spin in $S_n$ can be
chosen to be a primary field and that its OPE with itself is
just that for \W$(2n-1)$.  This connection between commuting charges
and \W-algebras will be further discussed in section 4.

Of course, all the above deliberations were based on
computer results involving currents of spin 13 and less.  We
now summarize the above and in the next sections prove by
analytic means some of the statements.

We have amassed sufficient evidence to conclude that:

(A) There exists an infinite number of sets of
mutually commuting operators constructed from the Kac-Moody
generators $J(z)$.  The series $S_n$ has a unique current at
every even spin and unique odd spin currents at spins
$1+m (h-1) $ for $h=2n-1$ and $m=0,1,2,\ldots$.

(B) The even spin currents can be built in terms of
the energy-momentum tensor $T={1\over 2} J^{\,2} + 3{(h-1)\over \sqrt{h+1}}
J'$ where the central charge is $c = {1\over h+1}(-3h^2 +
7h -2)$.

(C) The series $S_n$ can be associated with the
algebra \W$(2n-1)$, the current $p_{2n-1}$ being a primary field which
is the generator of \W$(2n-1)$.  All odd spin currents in $S_n$ are descendants
of $p_{2n-1}$.

One might ask how the above sets of commuting quantities compare with those
that are known from classical integrable systems.  For the $n$'th KdV hierarchy
there are conserved currents with spins 2, 3, $\ldots$, $n+1$ mod $n+1$, so
that the conserved charges have spins 1, 2, $\ldots$, n mod $n+1$.  These are
the exponents of the Lie algebra $sl(n)$ repeated modulo the Coxeter number,
and in fact for any Lie algebra $\cal G$ there exists an integrable system for
which the charges have spins equal to the exponents of $\cal G$ \cite{Wi,DS}.
These do not correspond to the spins of the currents for any of the series
found above.

\sect{Commuting currents with a $J^{\,3}$}

In this section we study in more detail the first series of currents,
$S_2$, containing a current $J^{\,3}$.  Some features of this series will
generalize to the other series, although there are aspects of this
series that will have no analogues for the others.

Let us first show that this series, which is defined by the existence of
commuting currents\footnote{We will loosely say that two currents commute
if their corresponding charges commute} of spins 3 and 4, corresponds to
the central charge having the value $c=-2$.  If we take an arbitrary
current $p_\ii$ of spin $\ii$, defined modulo derivatives,
\be
p_\ii = J^{\,\ii} + g(\ii) J^{\,\ii-4}(J')^2 + \ldots,
\en
it is straightforward to calculate its OPE with $J^{\,3}$.  We find that,
up to
derivatives, the coefficient of the single pole is given by
\be
{\ii(\ii-1)(\ii-2)(\ii-3)\over 4}J^{\,\ii-4} (J')^3 + 6 g(\ii)
J^{\,\ii-4}(J')^3+\ldots,
\en
so we must have
\be
g(\ii) = - {\ii \choose 4}
\en
in accordance with the computer results of the previous section.
For $\ii=4$ we find the current is $J^{\,4}-(J')^2$, and comparing with
(\ref{TJ}) we find that $c=-2$.  Hence we conclude that a spin-3 and a
spin-4 current can commute only if $c=-2$.

{}From the computer results of the previous section it is apparent that for
$c=-2$ or $x=0$ there is a considerable simplification in the form of the
currents.  In fact they are consistent with the formula
\be
p_\ii = \,:\!e^{-\phi}\partial^\ii e^\phi\!:\quad,
\label{p_\ii,c=-2}
\en
up to derivatives.
The generating function for this series is
\be
:\!e^{-\phi(z)+\phi(z+\alpha)}\!:\,=\sum_{r=0}^\infty {\alpha^r\over
r!}p_r(z).
\en
We can use this to show that integrals of the currents $p_\ii$ commute---to
this end we consider
\bea
I&\equiv&:\!e^{-\phi(z)+\phi(z+\alpha)}\!:\>:\!e^{-
\phi(w)+\phi(w+\beta)}\!:
\nonumber \\
& = &{(z-w)(z+\alpha-w-\beta)\over(z-w-\beta)(z+\alpha-w)}
:\!\exp\left(-\phi(z)+\phi(z+\alpha)-\phi(w)+\phi(w+\beta)\right)\!:.
\ena
Integrating this gives
\bea
\oint_0 dz\oint_z dw \,I &=& \nonumber \\[1mm]
&& \hspace{-3cm}{\alpha\beta\over \alpha+\beta}
\oint_0dz\,\{:\!\exp\left(-\phi(z-\beta)+\phi(z+\alpha)\right)\!:-
:\!\exp\left(-\phi(z)+\phi(z+\alpha+\beta)\right)\!:\},
\ena
since $\alpha$ and $\beta$ are small and so contained within the $w$
contour around $z$.  This can be seen to vanish by a shift of the $z$
integration contour, and as a consequence we can conclude that
\be
[Q_r,Q_s] \equiv \oint_0dz\oint_zdw \,p_r(z)p_s(w) = 0,\quad
\forall r,s.
\en

These results for the series $S_2$ can understood in terms of
the algebra \W$_\infty$ of reference \cite{PRS1,PRS2}.  This is a linear
algebra containing a quasiprimary field $V^i(z)$ with spin $i+2$ for
$i=0,1,2,\ldots$.  Examining the commutation relations of \W$_\infty$, as
given for example in eqn(3.2) of \cite{Pope}, it can be seen that the modes
$V^i_{-i-1}$ form an infinite set of commuting operators, which we might
think of as a Cartan subalgebra.  Since $V^0(z) = T(z)$, we recognize
$V^0_{-1}$ as $L_{-1}$, and similarly $V^1_{-2}$ is the mode $W_{-2}$ of
the spin-3 primary field $W(z)$.

For the case of $c=-2$, however, the \W$_\infty$ algebra has a realization
in terms of a complex fermion, and this can be bosonized to give a
realization in terms of a single real scalar field.  In terms of this
scalar field $\phi$, the field $V^i$ is proportional to
$:\!e^{-\phi}\partial^{i+2}e^\phi\!:$, up to derivatives.  We recognize
this as
the current given in eqn (\ref{p_\ii,c=-2}).  We also note that for
$c=-2$ the fields $V^i$ for $i\ge2$ can be written as composites of the
stress tensor $T$ and the spin-3 primary field $W$, so that the enveloping
algebra of the Zamolodchikov \W$(3)$-algebra
contains \W$_\infty$ as a linear subalgebra for $c=-2$.  This thus gives a
nice explanation of the presence of the infinitely many commuting
quantities in the enveloping algebra of \W$(3)$ in terms of a linear
algebra.

Let us summarize the main results of this section.  Starting from a \mbox{\sf
U}$(1)$
Kac-Moody algebra, we demanded a set of commuting currents that included
currents of spins 3 and 4.  We found that we could supplement these two
currents by an infinite number of other currents, one at each spin, that
formed a mutually commuting set.  These could be expressed in terms of a
stress tensor $T$ and a spin-3 primary field $W$ alone, where $T$ has a
central charge $c=-2$.  Furthermore these currents can be identified with a
Cartan subalgebra of \W$_\infty$, which in turn can be written in terms of
$T$ and $W$.  We therefore see that in terms of the algebra \W$_\infty$ it is
possible to understand very easily the existence of the commuting charges.
Although we do not know of a generalization of \W$_\infty$ that enables us to
gain a similar understanding of all the series of commuting charges, we shall
see in the next section that, for each series, there is a relation between the
odd-spin currents that allows to us to give a simple proof of the commutativity
of the corresponding charges.

\sect{Analytic results}
In this section we wish to give analytic derivations of some of the results
that we have obtained earlier.  More specifically, we shall give explicit
formulae for all of the odd spin fields in each of the series $S_n$, and we
shall prove that these fields commute with each other and with all of the
even spin fields in the series\footnote{Here, as before, we refer to fields
commuting when we really mean that their integrals commute.}.  We shall
also spell out in detail the connection of the series $S_n$ to a
\W-algebra.

Let us consider the forms of the fields of odd spin in $S_n$.  The series
$S_n$ exists for $c$ having the value $13-6n-
6/n$, with $n$ a positive integer, and the first
non-trivial field of odd spin has dimension $2n-1$ and can be
chosen to be primary.  From the work of Zamolodchikov \cite{Z} we expect there
to be
only two primary fields that commute with the even spin fields built out of
the stress tensor, namely the two fields with null descendants at level 3.
If we write an arbitrary value of the central charge as $c=13-6t-6/t$, the
two corresponding fields will have dimensions $h_{(3,1)}=2t-1$ and
$h_{(1,3)}=2/t-1$.  These two values of the dimension $h$ are related to
$c$ by
\bea
c&=&13-3(h+1)-{12\over h+1}\nonumber\\
&=& {1\over h+1}(-3h^2+7h-2).
\ena
Taking $t$ to be a positive integer $n$ we indeed expect to find a field of
dimension $2n-1$ that commutes with the commuting charges constructed from
$T$.  In order to write down an explicit
expression for this field in terms of the currents $J$
alone, we first consider the two fields of this dimension that can be
written as exponentials of $\phi$, namely
\be
V_{-\alpha_+} \equiv e^{ -\alpha_+ \phi}
\en
and
\be
V_{2\alpha_+ + \alpha_-} \equiv e^{(2\alpha_+ + \alpha_-)\phi},
\en
where $\alpha_+ = \sqrt{2n}$ and $\alpha_- = -\sqrt{2/n}$.
Each of these fields has a null descendant at level 3, obtained by acting
with the operator
\be
{\cal O}_{(3,1)} \,\equiv \, L_{-3} - {2\over 2n+1}L_{-1}L_{-2} +
{1\over2n(2n+1)}L_{-1}^3.
\en
It is only for $V_{-\alpha_+}$, however, that this null descendant is
actually zero, and so it is only in this case that the commutator of the
corresponding charges will actually vanish, rather than simply giving zero
in correlation functions.  We shall therefore concentrate our attention on
this field in what follows.

When $c=13-6n-6/n$, we can use the screening charge
$Q_+ \equiv \oint \exp \alpha_+ \phi$ to construct another primary field of
weight $2n-1$ that is annihilated by ${\cal O}_{(3,1)}$.  The commutator of
a Virasoro generator $L_n$ with $\exp \alpha_+ \phi$ is a total derivative,
so provided this field is single-valued the commutator of $Q_+$ with a
primary field will be another primary field of the same weight.  In general
this new primary
field could be zero, but that will not be the case here.  For the
values of $c$ in which we are interested, $\exp \alpha_+ \phi$ is indeed
local with respect to $\exp - \alpha_+ \phi$, and we can therefore
construct
another primary field of weight $2n-1$ by
\be
p_{2n-1}\equiv [Q_+,\exp - \alpha_+ \phi(z)] =
\oint_z dw \exp(\alpha_+ \phi(w))\exp(-\alpha_+ \phi(z)).
\en
This field has vanishing background charge and so is expressible entirely
in terms of the current $J$ and its
derivatives, and in fact it is given explicitly by
the term of order $\partial^0$ in the differential operator
\be
(\partial + \alpha_+ J)^{2n-1},
\en
in which the derivatives are taken to act on everything that occurs to the
right of them.
We claim that this is the first non-trivial field of odd spin occuring in
$S_n$. For the first few series this can be checked explicitly against the
computer results given in Section 2. To show that it is true in general, we
need to check that the charge constructed from this field
commutes with all of the even spin charges, which correspond to those that
occur in the quantum KdV equation. This
is a consequence of the fact that the screening charge
$\exp \alpha_+ \phi$ commutes with
any polynomial in $T$ and its derivatives, and in particular with the even spin
charges.  Since these even spin charges are even under
$\phi \rightarrow - \phi$, it follows that $\oint \exp - \alpha_+\phi$ also
commutes with these charges, as explained in ref \cite{SY}.

The above considerations immediately suggest a way to
construct infinitely many odd spin currents whose integrals  commute with the
even spin charges.
We take $Q_+$, as defined earlier, to be $\oint \exp \alpha_+ \phi$, and we
define also $Q_- = \oint \exp - \alpha_+ \phi$.  These two charges have
conformal weights $0$ and $2n-2$ respectively, and each commutes with all
of the even spin charges.
We then define an operator $\Delta$ that acts on an arbitrary field
$\Phi$ by
\bea
\Delta \Phi(z) &=& [Q_+,[Q_-,\Phi(z)]]\nonumber\\
&=& \oint_z dy \oint_{y,z} dx :\!e^{\alpha_+ \phi(x)}\!:\,:\!e^{-\alpha_+
\phi(y)}\!: \Phi(z).
\ena
If the integral of $\Phi$ commutes with the even spin charges, it is clear that
the integral of $\Delta \Phi$ will also commute with these charges.
Our  strategy is then to apply $\Delta$ repeatedly to the primary field
constructed above.  In fact it is interesting to note that the primary
field $p_{2n-1}$ can itself be written as
\be
p_{2n-1} = \Delta J,
\en
since $[Q_-, J(z)]=\exp-\alpha_+\phi(z)$.
If we write $\Phi^{(m)}$ for the $m$'th odd spin field in $S_n$, and $Q^{(m)}$
for the corresponding charge, we have
\be
\Phi^{(m)} = \Delta \Phi^{(m-1)},\quad \Phi^{(0)}=J,
\en
with similar formulae holding for $Q^{(m)}$.
Thus $\Phi^{(m)}$ is given by the multiple commutator
\be
\Phi^{(m)} = \Delta^m J =
\underbrace{[Q_+,[Q_-,\ldots[Q_+,[Q_-}_{m\rm\;times},J],
\en
which is given explicitly by
\be
\oint dx_m\oint dy_m,\ldots\oint dx_1\oint dy_1\,\{\prod_{i=1}^m
e^{\alpha_+\phi(z+x_i)}e^{-\alpha_+\phi(z+y_i)}\} J(z)
\en
where the integration contours satisfy $|x_m|>|y_m|\ldots|x_1|>|y_1|>0$.
Furthermore it is readily
seen to be expressible as a sum of terms, with each term being a product of
derivatives of exponentials of $\phi$, with the total \mbox{\sf U}$(1)$ charge
for each term in the sum being zero.  Thus $\Phi^{(m)}$ can be written in terms
of $J$ alone, with no exponentials of
$\phi$.  While it is conceivable that this multiple commutator could vanish,
and indeed this will be the case for some orderings of the $Q_+$ and
$Q_-$, we believe this will not happen in general for the ordering we have
chosen.  Let us demonstrate this for the first series, $S_2$.
To do this we take the generating functional
\bea
G[\beta] &=& \oint dz\, e^{-\phi(z) + \phi(z+\beta)},\nonumber\\
&=& \sum_n {\beta^n\over n!} Q^{(n)}
\ena
where the integration contour surrounds both the origin and $\beta$, and then
form
\be
\Delta G[\beta] \equiv \oint_0 dy \oint_y dx\,
e^{-2\phi(x)}
e^{2\phi(y)} G[\beta]
\en
which equals
\be
\oint_0 dz \oint_z dy \oint_{y,z} dx\,
{(x-z)^2(y-z-\beta)^2\over(x-y)^4(x-z-\beta)^2(y-z)^2}
\exp(-2\phi(x)+2\phi(y)-\phi(z)+\phi(z+\beta)).
\en
Carrying out the $y$ integration we obtain
\be
-2\beta{\partial\over\partial\beta}\oint\oint dx\,dz\,
{\beta\over(x-z)^2(x-z-\beta)^2} \exp(-2\phi(x)+\phi(z)+\phi(z+\beta)).
\en
We can now do the $x$ integration, giving
\be
4[{2\over\beta^2}-{2\over\beta}{\partial\over\partial\beta}+
{\partial^2\over\partial\beta^2}]\oint dz\,
\{\exp(-\phi(z)+\phi(z+\beta)) - \exp(\phi(z) - \phi(z+\beta))\}.
\en
This then leads to
\be
\Delta Q^{(n)} = \cases{0,&$n$ even;\cr {8n(n+1)\over
n!}Q^{(n+2)},&otherwise.\cr}
\en
Hence we can indeed construct an infinite sequence of non-zero charges as
multiple commutators in this way for $c=-2$, and as far as we are able to
calculate in the higher series we obtain non-zero charges in these cases also.

Having seen how to write explicit formulae for an infinite set of charges
commuting with the even spin charges, our next task is to understand why these
charges commute amongst themselves.  We start from the recursive definition for
the $m$'th
charge $Q^{(m)}$,
\bea
Q^{(1)}& =& [Q_+,Q_-] \nonumber \\
Q^{(m+1)} &=&[Q_+,[Q_-,Q^{(m)}]].\nonumber
\ena
We shall prove that $[Q^{(n)},Q^{(m)}] = 0$ for general $m$ and $n$, but we
begin by proving this at lowest order.  A crucial identity both in this simple
case and in general is that
\be
[Q_+,[Q_+,[Q_+,Q_-]]] = 0,
\label{Q_+^3}
\en
and similarly
\be
[Q_-,[Q_-,[Q_-,Q_+]]] = 0.
\label{Q_-^3}
\en
These relations can be obtained very simply just from looking at the
contour integrals that need to be done in order to evaluate these
expressions.
An essentially equivalent derivation for the first of these two relations
follows from the observation that
$[Q_+,[Q_+,[Q_+,\exp(-\alpha_+\phi)]]]$ has the same conformal weight as
$\exp(-\alpha_+\phi)$, namely
$2n-1$, but that it must be a descendant of $\exp2\alpha_+\phi$, which has
weight $2n+2$.  Evidently this is possible only if
$[Q_+,[Q_+,[Q_+,\exp(-\alpha_+\phi)]]]$ vanishes.
We now explore some of the consequences of the relations
(\ref{Q_+^3}) and (\ref{Q_-^3}), which we choose to rewrite in the form
\be
[Q_+,[Q_+,Q^{(1)}]]=0,\quad[Q_-,[Q_-,Q^{(1)}]]=0.
\en
It will be convenient to introduce the notation $Q_+X$ for the commutator
$[Q_+,X]$, where $X$ is an arbitrary field, and similarly for $Q_-$.  The
above relations then become
\be
Q_+^2Q^{(1)}=0,\quad Q_-^2 Q^{(1)} = 0.
\en
Let us start from
\be
Q^{(2)} = [Q_+,[Q_-,Q^{(1)}]]=[Q_-,[Q_+,Q^{(1)}]],
\en
which we write as
\be
Q^{(2)} = Q_+Q_-Q^{(1)}=Q_-Q_+Q^{(1)};
\en
the equality of these two expressions follows from the Jacobi identity.
Taking the commutator of these relations with $Q_+$ and $Q_-$ implies,
using the Jacobi identity again, that
\be
Q_+Q^{(2)} = [Q^{(1)},Q_+Q^{(1)}],\quad
Q_-Q^{(2)} =- [Q^{(1)},Q_-Q^{(1)}],
\en
and so
\bea
Q_+Q_-Q^{(2)}& = & -[Q_+Q^{(1)},Q_-Q^{(1)}]
-[Q^{(1)},Q^{(2)}]\nonumber \\[1mm]
Q_-Q_+Q^{(2)}& = & [Q_-Q^{(1)},Q_+Q^{(1)}] + [Q^{(1)},Q^{(2)}].
\nonumber
\ena
This implies that
\be
Q_+Q_-Q^{(2)} - Q_-Q_+Q^{(2)} = -2 [Q^{(1)},Q^{(2)}],
\en
whereas the Jacobi identity implies
\be
Q_+Q_-Q^{(2)} - Q_-Q_+Q^{(2)} = [Q^{(1)},Q^{(2)}].
\en
{}From this we conclude that
\be
[Q^{(1)},Q^{(2)}] = 0,
\en
and hence
\be
Q^{(3)}=Q_+Q_-Q^{(2)}=[Q_-Q^{(1)},Q_+Q^{(1)}].
\en
We have also that
\be
Q_+^2Q^{(2)} = Q_-^2Q^{(2)} = 0.
\en
This begins to suggest the following pattern:
\bea
[Q^{(i)},Q^{(j)}] & = & 0, \qquad\hbox{for}\quad i + j = n,\nonumber\\[1mm]
Q^{(n)} & \equiv & Q_+Q_-Q^{(n-1)}\nonumber\\[1mm]
        & = & [Q_-Q^{(k)},Q_+Q^{(l)}], \qquad\hbox{for}\quad k+l=n-1,
\nonumber\\[1mm]
Q_+^2Q^{(n-1)} & = & Q_-^2Q^{(n-1)} = 0.
\label{hypothesis}
\ena
We now prove this by induction.  We have seen that (\ref{hypothesis})
holds for $n=3$, so now assume that it is true for $n\le N$, for some $N$.
Then
\bea
Q_+^2Q^{(N)} & = &
Q_+^2[Q_-Q^{(i)},Q_+Q^{(j)}],\qquad
\hbox{for any}\quad i+j=N-1\nonumber\\[1mm]
    &=& Q_+[Q^{(i+1)},Q_+Q^{(j)}]\nonumber\\[1mm]
    &=& [Q_+Q^{(i+1)},Q_+Q^{(j)}]\nonumber\\[1mm]
    &=& {1\over 2} Q_+^2[Q^{(i+1)},Q^{(j)}] = 0,
\ena
and similarly we can prove that $Q_-^2Q^{(N)} = 0$.
Let us now look at $Q_+Q_-Q^{(N)}$ and
$Q_-Q_+Q^{(N)}$.  We have
\bea
Q_+Q_-Q^{(N)} &=&
Q_+Q_-[Q_-Q^{(i)},Q_+Q^{(j)}]\qquad
\hbox{for any}\quad i+j=N-1\nonumber\\[1mm]
&=&Q_+[Q_-Q^{(i)},Q^{(j+1)}]\nonumber\\[1mm]
&=&[Q^{(i+1)},Q^{(j+1)}] + [Q_-Q^{(i)},Q_+Q^{(j+1)}]
\ena
and
\bea
Q_-Q_+Q^{(N)} &=&
Q_-Q_+[Q_-Q^{(j)},Q_+Q^{(i)}]\nonumber\\[1mm]
&=&Q_-[Q^{(j+1)},Q_+Q^{(i)}]\nonumber\\[1mm]
&=&[Q^{(j+1)},Q^{(i+1)}] + [Q_-Q^{(j+1)},Q_+Q^{(i)}].
\ena
Hence
\bea
[Q^{(1)},Q^{(N)}] & = &
Q_+Q_-Q^{(N)} - Q_-Q_+Q^{(N)}\nonumber\\
 &=& 2 [Q^{(i+1)},Q^{(j+1)}] + [Q_-Q^{(i)},Q_+Q^{(j+1)}] + [Q_+Q^{(i)},Q_-
Q^{(j+1)}]\nonumber\\[1mm]
 &=& 2 [Q^{(i+1)},Q^{(j+1)}]+  Q_+Q_-[Q^{(i)},Q^{(j+1)}] -
[Q^{(i+1)},Q^{(j+1)}] - [Q^{(i)},Q^{(j+2)}]\nonumber\\[1mm]
 &=& [Q^{(i+1)},Q^{(j+1)}] - [Q^{(i)},Q^{(j+2)}].
\ena
This final expression can be solved for $[Q^{(k)},Q^{(N+1-k)}]$ in terms of
$[Q^{(1)},Q^{(N)}]$, giving
\be
[Q^{(k)},Q^{(N+1-k)}] = k [Q^{(1)},Q^{(N)}].
\en
In particular this implies $[Q^{(N)},Q^{(1)}] = N [Q^{(1)},Q^{(N)}]$, and
so $[Q^{(1)},Q^{(N)}]$ must vanish.  This in
turn implies that  $[Q^{(k)},Q^{(N+1-k)}]$ is zero.  It then follows that
$Q^{(N+1)} = [Q_-Q^{(i)},Q_+Q^{(j)}]$ for
any $i$ and $j$ such that $i+j=N$.  This completes the induction process.  We
conclude that $[Q^{(i)},Q^{(j)}]=0$ for all $i$ and $j$.

It is interesting to ask to what perturbation of a conformal field theory with
$c=13-6n-6/n$ the commuting charges found in this paper correspond.  This
amounts to finding an operator that commutes with these charges.  The charges
constructed from the even spin currents are differential polynomials in $T$ and
so will commute with the screening charges $\oint\exp\alpha_\pm\phi$. They also
commute with $\oint\exp-\alpha_\pm\phi$, on account of being even polynomials
in $\phi$.  The charges for the odd spin currents, however, are not constructed
from $T$ alone, and so will not commute with the above charges in general.
Nevertheless, since they are constructed from $\oint\exp\alpha_+\phi$ and
$\oint-\exp\alpha_+\phi$, and since $\alpha_+\alpha_- = -2$ implies that both
of these charges commute with $\oint\exp\pm\alpha_-\phi$, all the charges we
have found commute with $\oint\exp\alpha_-\phi$ and $\oint\exp-\alpha_-\phi$.
While the former is the screening charge, the latter is the field
$\phi_{(1,3)}$, which has weight $2/n-1$.  Thus the commuting charges are
conserved in the presence of the $\phi_{(1,3)}$ perturbation.  These remarks
are in agreement with those of Eguchi and Yang \cite{EY2}, who considered the
quantum sine-Gordon theory with Hamiltonian $\oint e^{\alpha_-\phi} -
e^{-\alpha_-\phi}$.  They observed that for particular values of the
sine-Gordon coupling constant extra odd-spin conserved currents existed for
spins $2n-1$ mod $(2n-2)$.  The detailed form of their charges is different
from ours,
however, and it is not possible to write them in terms of $J$ alone. Their
first
charge, for example, is given in our notation by $Q_+-Q_-$.

We now turn to the connection between the series of commuting charges and
certain \W-algebras.  We explained in the previous section that the first
series of charges was related to Zamolodchikov's \W$(3)$-algebra with the
central charge taking the value $c=-2$, and that for this value of $c$ the
enveloping algebra of the \W$(3)$-algebra contains a linear subalgebra which
is just \W$_\infty$.  In order to make a connection between \W-algebras and
the higher series, there was another aspect of the algebras that played an
important role in our proof of the commutativity of the charges constructed
from the odd-spin currents.  This was the fact that we had not just a
single primary field of conformal weight $2n-1$ for $c=13-6n-6/n$, but in
fact we made use of three distinct primary fields having this dimension.
In our free-field representation, these were $\exp (-\alpha_+ \phi)$, $Q_+
\exp (-\alpha_+ \phi)$ and $Q_+^2 \exp (-\alpha_+ \phi)$.  It is known that
for the values of $c$ that we are considering there exist \W-algebras
generated by the stress tensor and three primary fields each having spin
$2n-1$ \cite{K}. Let us denote these algebras by \W$((2n-1)^3)$. If we denote
the primary fields by $W^i(z)$, for $i=1$,2 and 3, they have operator
product expansions of the form
\be
[W^i][W^j] = \delta^{ij}[I] + \epsilon^{ijk}[W^k],
\en
so that there is an \mbox{\sf SU}$(2)$-like structure present.  The operator
$Q_+$ can be considered as an \mbox{\sf SU}$(2)$ raising operator.  We have
seen that acting repeatedly with $Q_-$ on the multiplet of primary fields gives
other spin-1 \mbox{\sf SU}$(2)$ multiplets of higher conformal weight.  The
fields in these higher multiplets are no longer primary, but their integrals
give rise to the infinite sets of commuting charges we have found.

\sect{Discussion}
One of the most interesting aspects of the results we have found is the
presence of higher symmetry algebras underlying the series of commuting
charges, namely \W$((2n-1)^3)$ for the series $S_n$.This symmetry played an
important role in our proof of the commutativity of the odd spin currents, and
in a sense it controlled the structure of the commuting set.

For the first series of charges, we found that the \W$(3)$ algebra could be
extended to \W$_\infty$.  The commuting quantities could then be thought of as
some sort of Cartan subalgebra of \W$_\infty$.  One may hope that for the
higher series $S_n$, the algebra \W$(2n-1)$ will have a similar extension to an
infinite-dimensional linear algebra and the same phenomenon occur.

One might also expect other groups to play a role in understanding the
commuting currents.  A large class of solutions of the KP hierarchy and its
reductions have been formulated using the Grassmanian approach of Sato and of
Segal and Wilson---see for example \cite{DKJM,SW}.  The KdV hierarchy is
associated to the group affine \mbox{\sf U}$(2)$, and the Grassmanian is the
coset space of this group divided by \mbox{\sf U}$(2)$.  To each point of the
Grassmanian is associated the Baker function, and the logarithm of this
function contains as coefficients in its power series all of the conserved
quantities. Furthermore it is possible to view the commuting charges as the
elements $J^3_n$, $n>0$, of an $\widehat{\mbox{\sf SU}(2)}$ Kac-Moody algebra.
It is natural to ask how one might generalize the construction of conserved
quantities via the Baker function to the quantum case, and to suppose that a
quantum analogue of affine \mbox{\sf U}$(2)$ will play a role.

The general form of the even spin currents has so far not been found explicitly
except for $c=-2$.  However, the form of the currents in terms of $J$ given in
eqn (\ref{fields}) strongly suggests that a closed form expression can be
found.  Such an expression would be likely to involve the factors
$\exp{\alpha_{\pm}\phi}$ and $\exp{-\alpha_{\pm}\phi}$.  Indeed, one way to
obtain quantities commuting with $\exp\pm\alpha_-\phi$ is to exploit the
identity $(\exp\pm\alpha_-\phi)^n=0$ \cite{BMcCP}, where $\alpha_-^2=2/n$, as a
consequence of which $(\exp\pm\alpha_-\phi)^{n-1} X$ will commute with
$\exp\pm\alpha_-\phi$ for any $X$.  For the first series $n=2$, and it can be
checked that in this case the currents can be written as $\oint dw
:e^{\alpha_-\phi(w)}:
:e^{\alpha_+\phi(z)}\partial^m e^{\alpha_-\phi(z)}:$.  Unfortunately we do not
know how to generalise this formula to the higher series.

Acknowledgement: M.D. Freeman and K. Hornfeck are grateful to the UK Science
and Engineering Research Council for financial support.
%
%
%

\end{document}